\documentstyle[12pt]{article} 
  \topmargin - 0.5 cm 
\begin{document}

\centerline{\bf Fractional supersymmetric Quantum Mechanics} 
\centerline{\bf as a set of replicas of} 
\centerline{\bf ordinary   supersymmetric Quantum Mechanics\footnote{Accepted 
for publication in Physics Letters A.}}

\bigskip
\bigskip

\centerline{M.~Daoud\footnote{Permanent address: 
Laboratoire de Physique de la Mati\`ere Condens\'ee, 
Facult\'e des Sciences, Universit\'e Ibn Zohr, 
BP 28/S, Agadir, Morocco.}, M.~Kibler\footnote{Correspondence to:
M. Kibler; e-mail: m.kibler@ipnl.in2p3.fr.}}

\medskip
\medskip

\centerline{Institut de Physique Nucl\'eaire de Lyon,}

\centerline{IN2P3-CNRS et 
Universit\'e Claude Bernard,}

\centerline{43 Bd du 11 Novembre 1918, 
F-69622 Villeurbanne Cedex, France}

\bigskip
\bigskip

\begin{abstract}
A connection between fractional
supersymmetric quantum mechanics
and ordinary supersymmetric quantum 
mechanics is established in this Letter.
\end{abstract}

\bigskip
\bigskip

0. Although {\it ordinary} supersymmetric Quantum 
Mechanics (sQM) was introduced more than 20 years 
ago, its extension as {\it fractional} sQM is still 
the object of numerous works. The parentage between 
ordinary sQM and fractional sQM needs to be clarified.
In particular, we may ask the question: Can fractional
sQM be reduced to ordinary sQM as far as spectral analyses
are concerned? 
It is the aim of this work to study a connection
between fractional sQM of 
order $k$ and ordinary sQM  
corresponding to $k=2$. We consider here the case where the number 
of supercharges is 
equal to 1 (corresponding to 2 supercharges related via Hermitean
conjugation).

\smallskip
1. Our definition of {\it fractional} sQM of order $k$, with
$k \in {\bf N} \setminus \{0,1\}$, is as follows. Following Refs.~[1-4], 
a doublet of linear operators $(H,Q)_k$, with $H$ a self-adjoint operator and
$Q$ a supersymmetry operator, acting on a separable Hilbert space and satisfying 
the relations 
$$
Q_- = Q,                             \quad 
Q_+ = Q^{\dagger}                    \quad 
(\Rightarrow \ Q_-^{\dagger} = Q_+), \quad 
Q_{\pm}^k = 0 
\eqno (1{\rm a})
$$
$$
Q_- ^{k-1} Q_+  +  Q_- ^{k-2} Q_+ Q_- 
                                      +   \cdots 
                                      +   Q_+ Q_- ^{k-1}
                                      =       Q_- ^{k-2} H
\eqno (1{\rm b})
$$
$$
[H , Q_{\pm}] = 0
\eqno (1{\rm c})
$$
is said to define a $k$-fractional supersymmetric 
quantum-mechanical system (see also Refs.~[5-8]). The operator $H$ is 
the Hamiltonian of the system spanned by the two 
(dependent) supercharge operators $Q_-$ and $Q_+$. In 
the special case $k=2$, the system described by 
a doublet of type $(H,Q)_2$ is referred to as  
an ordinary supersymmetric quantum-mechanical
system~; it corresponds to a $Z_2$-grading
with fermionic and bosonic states.  

\smallskip
2. We now introduce a generalized Weyl-Heisenberg 
algebra $W_k$, with $k \in {\bf N} \setminus \{0,1\}$, from 
which we can construct $k$-fractional supersymmetric 
quantum-mechanical systems. The algebra $W_k$ is spanned 
by four linear operators, {\it viz.},   
$X_-$ (annihilation operator), 
$X_+$ (creation operator), 
$N$ (number operator) and 
$K$ ($Z_k$-grading operator). The operators $X_-$ and $X_+$ are connected via
Hermitean conjugation; $N$ is a self-adjoint operator and $K$ is a 
unitary operator. The four operators satisfy the relationships
$$
 [X_- , X_+] = \sum_{s=0}^{k-1} f_s(N) \> \Pi_s, \
 [N , X_{\pm}] = {\pm} X_{\pm},                  \ 
 [K , X_{\pm}]_{q^{\pm 1}} = 0,                  \
 [K , N] = 0,                                    \
 K^k = 1
\eqno (2)
$$
Here, the functions $f_s : N \mapsto f_s(N)$ are arbitrary functions 
subjected to the constraints $f_s(N)^{\dagger} = f_s(N)$. Furthemore, the 
Hermitean operators $\Pi_{s}$ are defined by  
$$
\Pi_{s} = \frac{1}{k}  \>  \sum_{t=0}^{k-1}  \>  q^{-st} \> K^t
$$
where 
$$
q = \exp \left( \frac{2 \pi {\rm i}}{k} \right)
$$
is a root of unity, so that they are projection 
operators for the cyclic group $C_k$. Finally, 
$[K , X_{\pm}]_{q^{\pm 1}}$ stands for the deformed commutator 
$K X_{\pm} - q^{\pm 1} X_{\pm} K$.

\smallskip
3. The operators $X_-$, $X_+$ and $K$ can be 
realized in terms of $k$ pairs $(b(s)_- , b(s)_+)$
of deformed bosons with
$$
[ b(s)_- , b(s)_+ ] = f_s(N)
$$
and one pair $(f_- , f_+)$ of $k$-fermions with
$$
[ f_- , f_+ ]_q = 1, \quad f_{\pm}^k = 0
$$
The $f$'s commute with the $b$'s. Of course, we have 
$b(s)_+  = b(s)_-^{\dagger}$ but 
$f_+ \not=    f_-^{\dagger}$ except for $k = 2$. The 
$k$-fermions introduced in [9] and recently 
discussed in [10] are objects interpolating
between fermions and bosons (the case $k=2$ corresponds 
to ordinary fermions and the case $k \to \infty$ to ordinary
bosons); the $k$-fermions also share some features of the 
anyons introduced in [11,12]. For $k$ arbitrary in 
${\bf N} \setminus \{0,1\}$, the realization  
$$
K = [f_- , f_+ ]
$$
$$
X_- = 
\left( f_-   +    \frac{ f_+ ^{k-1}}{[k - 1]_q !} \right)
\sum_{s=0}^{k-1} b(s)_- \> \Pi_s
$$
$$
X_+ = 
\left( f_-   +    \frac{ f_+ ^{k-1}}{[k - 1]_q !} \right)^{k-1}
\sum_{s=0}^{k-1} b(s)_+ \> \Pi_s
$$
has been discussed in Ref.~[8]. Here, we have  
$[ n ]_q ! = 
 [ 1 ]_q \>
 [ 2 ]_q \> \cdots \> 
 [ n ]_q$ 
(with $[ 0 ]_q ! = 1$) and
the symbol $[ \ ]_q$ is defined by
$$
[n]_q = {1 - q^n \over 1 - q}
$$
where $n \in {\bf N}$.

\smallskip
4. An Hilbertean representation of $W_k$ can be constructed
in the following way. Let ${\cal F}$ be the Hilbert-Fock space 
on which the generators 
$X_-$, $X_+$, $N$ and $K$ act. Since $K$ 
is a cyclic operator of order $k$, the space ${\cal F}$ can be graded as
$$
 {\cal F} = \bigoplus_{s=0}^{k-1} {\cal F}_s
$$  
where the subspace 
${\cal F}_s = \{ |n , s \rangle : n = 1, 2, \cdots, d \}$ 
is a $d$-dimensional space ($d$ can be finite or infinite).
The representation is given by
$$
K | n , s \rangle = q^{s} | n , s \rangle, \quad
N | n , s \rangle = 
n | n , s \rangle
$$
$$
X_- | n , s   \rangle = \sqrt {F_{s}(n)} \> \cases{ 
                       | n-1 , s-1 \rangle \mbox{ if } s \not=0
\cr \cr
                       | n-1 , k-1 \rangle \mbox{ if } s     =0
                                               }
$$
$$
X_+ | n , s   \rangle = \sqrt {F_{s+1}(n+1)} \> \cases{ 
                       | n+1 , s+1 \rangle \mbox{ if } s \not=k-1
\cr \cr
                       | n+1 , 0   \rangle \mbox{ if } s     =k-1
                                                   }
$$
where the function $F$ is a structure function such that
$$
F_{s+1}(n+1) - F_{s}(n) = f_s(n)
\eqno (3)
$$
with $F_s(0) = 0$.

\smallskip
5. We are now in a position to associate a $k$-fractional 
supersymmetric quantum-mechanical system to the algebra
$W_k$ characterized by a given set of functions 
$\{ f_s : s = 0, 1, \cdots, k-1 \}$. We define the supercharge
$Q$ via
$$
Q           \equiv Q_- = X_- (1 - \Pi_{1}) \Leftrightarrow 
Q^{\dagger} \equiv Q_+ = X_+ (1 - \Pi_{0}) 
\eqno (4)
$$
There are $k$ equivalent definitions of $Q$ corresponding to
the $k$ circular permutations of $1, 2, \cdots, k-1$; our choice,
which is such that $Q | n,1 \rangle = 0$, is adapted to the sequence 
$H_k, H_{k-1}, \cdots, H_1$ to be considered below.
By making reapeated use of Eqs.~(1), (2) and (4),
we can derive the operator 
$$
H = (k-1) X_+ X_- -
\sum_{s=3}^k 
\sum_{t=2}^{s-1} (t-1) \> f_t(N-s+t) \> \Pi_s - 
\sum_{s=1}^{k-1} 
\sum_{t=s}^{k-1} (t-k) \> f_t(N-s+t) \> \Pi_s 
\eqno (5)
$$ 
which is self-adjoint and commutes 
with $Q_-$ and $Q_+$. 
(Equation (5) and some other relations below include $\Pi_k$. Indeed, in view of
the cyclic character of $K$, we have $\Pi_k = \Pi_0$ so that the action of terms
involving $\Pi_k$ is quite well-defined on the space ${\cal F}$.)
As a result, the doublet $(H, Q)_k$ associated to $W_k$
satisfies Eq.~(1) and thus defines a $k$-fractional 
supersymmetric quantum-mechanical system.

\smallskip
6. In order to establish a connection between {\it fractional}
sQM (of order $k$) and {\it ordinary} sQM (of order $k = 2$),
it is necessary to construct subsystems from the doublet $(H, Q)_k$ 
that correspond to ordinary supersymmetric quantum-mechanical systems.
This may be achieved in the following way. Equation (5) can be 
rewritten as 
$$
H = \sum_{s=1}^{k} H_{s} \> \Pi_{s}
\eqno (6) 
$$
where 
$$
H_s \equiv H_s(N) 
    = (k-1) F(N) - \sum_{t=2}^{k-1} (t-1) \> f_t(N-s+t)
    + (k-1)        \sum_{t=s}^{k-1}          f_t(N-s+t) 
\eqno (7)
$$
It can be shown that the operators 
$H_k \equiv H_0, H_{k-1}, \cdots, H_{1}$
turn out to be isospectral operators. 
By introducing 
$$
X(s)_- = 
  \sum_{n  }       [H_s(n)  ]^{\frac{1}{2}} | n-1 , s-1 \rangle \langle n , s   |
$$
$$
X(s)_+ = 
  \sum_{n  }       [H_s(n+1)]^{\frac{1}{2}} | n+1 , s   \rangle \langle n , s-1 |
$$
it is possible to factorize $H_s$ as 
$$
H_s = X(s)_+ \> X(s)_-
$$
modulo the omission of the ground state $|0,s\rangle$ 
  (which amounts to substract the corresponding eigenvalue 
  from the spectrum of $H_s$).
Let us now define: (i) the two (supercharge) operators
$$
q(s)_- = X(s)_- \> \Pi_s, \quad 
q(s)_+ = X(s)_+ \> \Pi_{s-1}
$$ 
and (ii) the (Hamiltonian) operator 
$$
h(s) = X(s)_- \> X(s)_+ \> \Pi_{s-1}   +   X(s)_+ \> X(s)_- \> \Pi_s
\eqno (8)
$$
It is then a simple matter of calculation to prove that 
$h(s)$ is self-adjoint and that
$$
q(s)_+ = q(s)_-^{\dagger},             \quad
q(s)_{\pm}^2 = 0,                      \quad 
h(s) = q(s)_- q(s)_+ + q(s)_+ q(s)_-,  \quad 
[ h(s) , q(s)_{\pm} ] = 0
$$
Consequently, the doublet $(h(s), q(s))_2$, with
$q(s) \equiv q(s)_-$, 
satisfies Eq.~(1) with $k=2$ and thus 
defines an ordinary sypersymmetric 
quantum-mechanical system (corresponding to $k=2$).  

\smallskip  
7. 
The Hamiltonian $h(s)$ is amenable to a form more 
appropriate for discussing the link between ordinary 
sQM and fractional sQM. Indeed, we can show that 
$$
            X(s)_- \> X(s)_+ = H_s(N+1)
\eqno (9)
$$
Then, by combining Eqs.~(2), (3), (7) and (9), 
Eq.~(8) leads to the important relation 
$$
h(s) = H_{s-1} \> \Pi_{s-1} + H_{s} \> \Pi_{s}
\eqno (10)
$$
to be compared with the expansion of $H$ in terms 
of supersymmetric partners $H_s$ (see Eq.~(6)). 

\smallskip  
8. 
To close this Letter, let us sum up the 
obtained results and offer some conclusions.

Starting from a $Z_k$-graded algebra $W_k$,
characacterized by a set $ \{ f_s : s = 0, 1, \cdots, k-1 \} $,
it was shown how to associate a $k$-fractional supersymmetric 
quantum-mechanical system $(H,Q)_k$ 
characterized by an Hamiltonian $H$
and a supercharge $Q$. 

The extended Weyl-Heisenberg algebra $W_k$ 
covers numerous algebras describing exactly solvable 
one-dimensional systems. The particular system 
corresponding to a given set 
$\{ f_s : s = 0, 1, \cdots, k-1 \}$ yields, in a Schr\"odinger
picture, a particular dynamical system with a specific potential. Let 
us mention two interesting cases.
The case
$$
\forall s \in \{ 0, 1, \cdots, k-1 \} \ : \ 
f_s(N) = f_s \mbox{ independent of } N
$$
corresponds to systems with cyclic shape-invariant potentials 
(in the sense of Ref.~[13]) 
and the case
$$
\forall s \in \{ 0, 1, \cdots, k-1 \} \ : \ 
f_s(N) = a N + b  \mbox{ where } (a,b) \in {\bf R}^2
$$
to systems with translational shape-invariant potentials 
(in the sense of Ref.~[14]). For instance, 
the case $(a = 0, b > 0)$ 
corresponds to the harmonic oscillator potential, 
the case $(a < 0, b > 0)$ to the Morse potential and
the case $(a > 0, b > 0)$ to the P\"oshl-Teller potential. For these 
various potentials, the part of $W_k$ spanned by $X_-$, $X_+$ and $N$ 
can be identified with the ordinary Weyl-Heisenberg algebra 
for $(a = 0, b \not= 0)$,
                  with the su(1,1) Lie algebra 
for $(a > 0, b > 0)$ and  
                  with the su(2)   Lie algebra 
for $(a < 0, b > 0)$. These matters shall be the subject 
of a forthcoming paper. 
 
The Hamiltonian $H$ for the system $(H,Q)_k$
was developed as a superposition of $k$ isospectral 
supersymmetric partners $H_0, H_1, \cdots, H_{k-1}$.

The system $(H,Q)_k$ itself, corresponding to $k$-fractional
sQM, was expressed in terms of $k-1$ sub-systems $(h(s),q(s))_2$, 
corresponding to ordinary sQM. The Hamiltonian $h(s)$ is given 
as a sum involving the supersymmetric partners 
$H_{s-1}$ and $H_s$ (see Eq.~(10)). Since the 
supercharges $q(s)_{\pm}$ commute with the Hamiltonian $h(s)$, 
it follows that 
$$
H_{s-1} X(s)_- = X(s)_- H_{s  }, \quad 
H_{s  } X(s)_+ = X(s)_+ H_{s-1}
\eqno (11)
$$
As a consequence, the operator 
          $X(s)_+$ 
(respect. $X(s)_-$) makes it 
possible to pass from the spectrum of $H_{s-1}$ (respect. $H_{s  }$) 
                        to the one of $H_{s}  $ (respect. $H_{s-1}$). This 
result is quite familiar for ordinary sQM (corresponding to $s=2$). Note 
that Eq.~(11) is reminiscent of the intertwining method based on the
Darboux transformation and on the factorization method which are 
useful for studying superintegrability of quantum systems.

For $k=2$, the operator $h(1)$ is nothing but the total Hamiltonian $H$ 
corresponding to ordinary sQM. For
arbitrary $k$, the other operators $h(s)$ are simple replicas (except for the
ground state of $h(s)$) of $h(1)$. It is in 
this sense that $k$-fractional sQM 
can be considered as a set of $ k-1 $ replicas of ordinary sQM 
typically described by $(h(s),q(s)_{\pm})_2$. Along this vein,
it is to be emphasized that 
$$
H = q(2)_- \>
    q(2)_+ + \sum_{s=2}^{k} 
    q(s)_+ \> 
    q(s)_-
$$
which can be identified to $h(2)$ for $k=2$.

Thanks are due to the referee for pertinent and constructive remarks.

\end{document}